\documentclass[letterpaper,english, twocolumn, aps]{revtex4-1}
\usepackage[T1]{fontenc}
\usepackage{textcomp}
\usepackage{amsmath}
\usepackage{amssymb}
\usepackage{graphicx}
\usepackage[francais]{babel}
\usepackage[utf8]{inputenc}
\usepackage[T1]{fontenc}
\usepackage{natbib,hyperref}
\usepackage{wasysym}
\makeatletter

\providecommand{\tabularnewline}{\\}

\makeatother

\usepackage{babel}
\begin{document}

\preprint{APS/123-ALP}

\title{Optimization of laser-plasma injector via beam loading effects using ionization-induced injection}

\author{P. Lee}
\email{patrick.lee@u-psud.fr}
\affiliation{LPGP, CNRS, Univ. Paris-Sud, Université Paris-Saclay, 91405, Orsay, France}

\author{G. Maynard}
\affiliation{LPGP, CNRS, Univ. Paris-Sud, Université Paris-Saclay, 91405, Orsay, France}

\author{T. L. Audet}
\affiliation{LPGP, CNRS, Univ. Paris-Sud, Université Paris-Saclay, 91405, Orsay, France}

\author{R. Lehe}
\affiliation{Lawrence Berkeley National Laboratory, Berkeley, CA 94720, USA}

\author{J.-L.Vay}
\affiliation{Lawrence Berkeley National Laboratory, Berkeley, CA 94720, USA}

\author{B. Cros}
\email{brigitte.cros@u-psud.fr}
\affiliation{LPGP, CNRS, Univ. Paris-Sud, Université Paris-Saclay, 91405, Orsay, France}

\date{\today}
\begin{abstract}
Simulations of ionization induced injection in a laser driven plasma wakefield show that high-quality electron injectors in the 50-200 MeV range can be achieved in a gas cell with a tailored density profile. Using the PIC code Warp with parameters close to existing experimental conditions, we show that the concentration of $\mathrm{N_2}$ in a hydrogen plasma with a tailored density profile is an efficient parameter to tune electron beam properties through the control of the interplay between beam loading effects and varying accelerating field in the density profile. For a given laser plasma configuration, with moderate normalized laser amplitude, $a_0=1.6$ and maximum electron plasma density, $n_{e0}=4\times 10^{18}\,\mathrm{cm^{-3}}$, the optimum concentration results in a robust configuration to generate electrons at 150~MeV with a rms energy spread of 4\% and a spectral charge density of 1.8~pC/MeV.
\end{abstract}

\pacs{33.15.Ta}

\keywords{Suggested keywords}

\maketitle

\section{\label{sec:level1}Introduction}
Laser Wakefield Acceleration (LWFA) relies on an underdense plasma to transfer the energy from a laser beam to a trailing bunch of electrons, either injected internally or externally. The ponderomotive force of a laser pulse moving through the plasma pushes electrons ahead of the pulse and to the sides, creating in its wake a periodic structure of rarefaction and concentration of electrons. This electronic density perturbation results in a plasma wave which is characterized by strong electric and magnetic fields, known as wakefields \cite{tajima_laser_1979,esarey_physics_2009,malka_laser_2012}, with  acceleration gradients larger than $100\,\mathrm{GV/m}$ \cite{leemans_gev_2006,nakamura_gev_2007}. These peak gradients make LWFA a promising option towards compact high energy electron accelerators with a wide range of applications.

Multistage acceleration schemes \cite{leemans_laser-driven_2009}, where the electron beam is extracted from one module and injected into the subsequent module for further acceleration, would provide scalability and control, and are actively investigated for the development of future high energy accelerators. In these schemes, the electron injector is expected to produce a high-quality electron beam with narrow energy spread and small emittance. Many efforts are devoted to study mechanisms able to control the electron beam properties of the injector based on LWFA \cite{geddes_high-quality_2004,mangles_monoenergetic_2004,faure_laserplasma_2004}.

Several mechanisms can be used for the injector. Among them are self-injection \cite{mangles_monoenergetic_2004,geddes_high-quality_2004,faure_laserplasma_2004,kalmykov_electron_2011}, density-transition based injection \cite{bulanov_laser_1997}, shock-front injection \cite{fourmaux_quasi-monoenergetic_2012,brijesh_tuning_2012,burza_laser_2013,buck_shock-front_2013,schmid_density-transition_2010},  colliding pulse injection \cite{esarey_electron_1997}, and  ionization induced injection schemes \cite{mcguffey_ionization_2010,pak_injection_2010,clayton_self-guided_2010}.
This last mechanism provides a large number of parameters to control the injection of electrons,  resulting in a larger number of accelerated electrons at comparatively low laser intensity, and can be combined with other mechanisms for improved control. This makes it a promising mechanism for injector optimization.
Ionization induced injection utilizes the large difference in ionization potentials between successive ionization states of trace of high $Z$ atoms, to create electrons at selected phases of the wakefield. The use of trace high $Z$ atoms brings about a high beam charge \cite{couperus_demonstration_2017}, an additional degree of freedom, allowing electron trapping at lower plasma densities, and use of lower laser intensities as compared to the self-injection scheme. However, this injection mechanism tends to produce electron beams with a large energy spread because injection occurs continuously as long as the laser intensity exceeds the ionization threshold in the volume of mixed gas length, or until some competing mechanism, like beam loading, stops injection.

In several experiments a few mm-long mixed gas volume is followed by a volume where pure gas is injected \cite{liu_all-optical_2011,pollock_demonstration_2011,gonsalves_tunable_2011,kim_enhancement_2013,wang_quasi-monoenergetic_2013,golovin_tunable_2015}, the second volume acting as an accelerator and energy filter; in these experiments, the electrons generated in the first volume have a large energy spread, implying that the mixed gas length is still longer than optimum and efficiency of coupling to the subsequent accelerating region can be improved. For this purpose, hybrid mechanisms including tailoring of gas-density profile \cite{zeng_controlled_2012,thaury_shock_2015} and using moderate power pulses \cite{kamperidis_low_2014} on top of ionization induced injection scheme were introduced to limit the injection length, and promising results were obtained.

A numerical investigation of the dynamics of ionized-induced injected electrons in the tailored profiles was previously reported in \cite{lee_dynamics_2016}. The study conducted in  \cite{lee_dynamics_2016} emphasizes on the influence of the shape of the density down ramp profiles on the generated electron beam properties. In this article, the study will focus on yet another control parameter in the ionization induced injection scheme, namely the concentration of trace atoms. As pointed out in \cite{chen_theory_2012}, using nitrogen as the trace atom, the charge and the energy spread both increase with the trace atom concentration, with a dramatic increase of energy spread when nitrogen concentration exceeds $3\%$.

In this article, we report on a detailed numerical study of the influence of nitrogen concentration on the electron beam properties, at moderate laser intensity and plasma density, for low concentration of traced atoms ($\leq2\%$). We demonstrate that at an optimal nitrogen concentration, the beam-loaded electric field induced by trapped electrons can be used to control the energy spread;  with $C_\mathrm{N_2} = 0.35\,\%$, an electron beam of rms energy spread $\Delta\mathcal{E}_{rms}/\left<\mathcal{E}\right>\approx4\%$ is obtained. The configuration with this optimal concentration is proven to be robust with respect to changes of other plasma properties such as the plasma density down ramp profile, and the extension of the plasma length.

The remaining of the paper is organized as follows. Section \ref{sec:numerical} explains the rationale behind the choice of parameters including the selection of density profile characteristics (Section \ref{sec:target}) of the in-house variable length gas cell, ELISA \cite{audet_electron_2016}, and summarizes the numerical setup for PIC simulations. The control of ionization induced injection using nitrogen concentration, and a moderate power laser pulse, propagating in a single-stage mixed-gas cell, is analyzed in Section \ref{sec:analysis}. Using the optimized nitrogen concentration and taking advantage of the beam loading effects, the electron beam energy can be tuned by extending the plasma length while preserving other properties, as presented in Section \ref{sec:extension}. The robustness of this range of optimum concentration regarding laser fluctuation is discussed in Section \ref{sec:laser}.

\section{Choice of laser-plasma parameters}
\label{sec:numerical}
\subsection{Regime of acceleration}

Laser-plasma parameters are chosen by taking into account the following criteria. The required energy range of the electron beam for an injector is between $\mathrm{50-200 \, MeV}$. The lower limit at $\mathrm{50 \, MeV}$ ensures that space charge effects will not be dominant, and that energy spread can be minimized as it scales as $1/\gamma^{2}$, where $\gamma = (1-(v/c)^2)^{-1/2}$, is the Lorentz factor, $v$ the electron velocity and $c$, the speed of light. The upper limit is fixed at $\mathrm{200\,MeV}$ to allow for a compact transport line for electron beam manipulation before coupling to the first accelerating structure. In this study, electron acceleration up to an energy of $150\,\mathrm{MeV}$ was chosen. In addition, the electron bunch is required to have a small normalized transverse emittance of $\varepsilon_n\sim 1\,\mathrm{mm\,mrad}$, a small energy spread (typically less than 10\%) and a large enough charge ($\mathrm{\geq}$ 10 $\mathrm{pC}$).

An a priori favorable configuration to produce a quasi-monoenergetic beam is to have the acceleration length $L_{acc}$ close to the electron dephasing length in order to optimize the phase space rotation, therefore at moderate intensities $L_{acc}\propto (\lambda_p^3/\lambda_0^2)\propto \max (n_{e0})^{-3/2}$, with $\lambda_p$ plasma wavelength, $\lambda_0$ laser wavelength, and $n_{e0}$ the maximum electron number density on axis. Considering all these factors and results from a previous study \cite{lee_dynamics_2016}, we set  $\max (n_{e0})$ to $4\times10^{18}\mathrm{cm}^{-3}$, with a target length of $\sim 1 \, \mathrm{mm}$.

At a given coordinate $z$ along the laser axis, the maximum value of laser amplitude in normalized units is defined by $a_0(z) = \max_{r,t}[eA(r,z,t)/m_e\omega c]$, where $\omega$ is the laser frequency, $e$ the electron charge, and $m_e$ the electron mass, and $A$ is the amplitude of the vector potential. The value of $a_0(z)$ at the focal plane in vacuum is chosen to be $1.6$ as it is large enough to ionize and inject the $6^{th}$ electron of nitrogen but not too large to provoke self-trapping of electrons after self-focusing in the plasma, the limit of which is $\sim 4$ \cite{pak_injection_2010}. The laser duration $\tau_L$ and waist $w_L$ are determined in such a way to efficiently excite the plasma wave using the resonance condition, $\tau_L\omega_p\sim 2\pi$. For $\max(n_{e0})=4\times 10^{18}$ and $a_0=1.6$, $\tau_L= 20\,\mathrm{fs}$ at full-width at half-maximum (FWHM), and $w_L= 16\,\mathrm{\mu m}$ at $1/e^2$ of the laser intensity. The laser pulse is assumed to have Gaussian temporal and spatial profiles. Apart from the bi-Gaussian profiles, the aforementioned laser parameters are close to the ones obtained in laser facilities equipped with a $>50\,\mathrm{TW}$ power laser. In accordance with previous studies \cite{audet_investigation_2016,lee_dynamics_2016}, the laser is focused at the exit of the target to delay the triggering of the ionization and injection of electrons so as to constrain the injection volume to limit the energy spread.

\subsection{Gas profile description}
\label{sec:target}

\begin{figure}[h]
\includegraphics[scale=0.25]{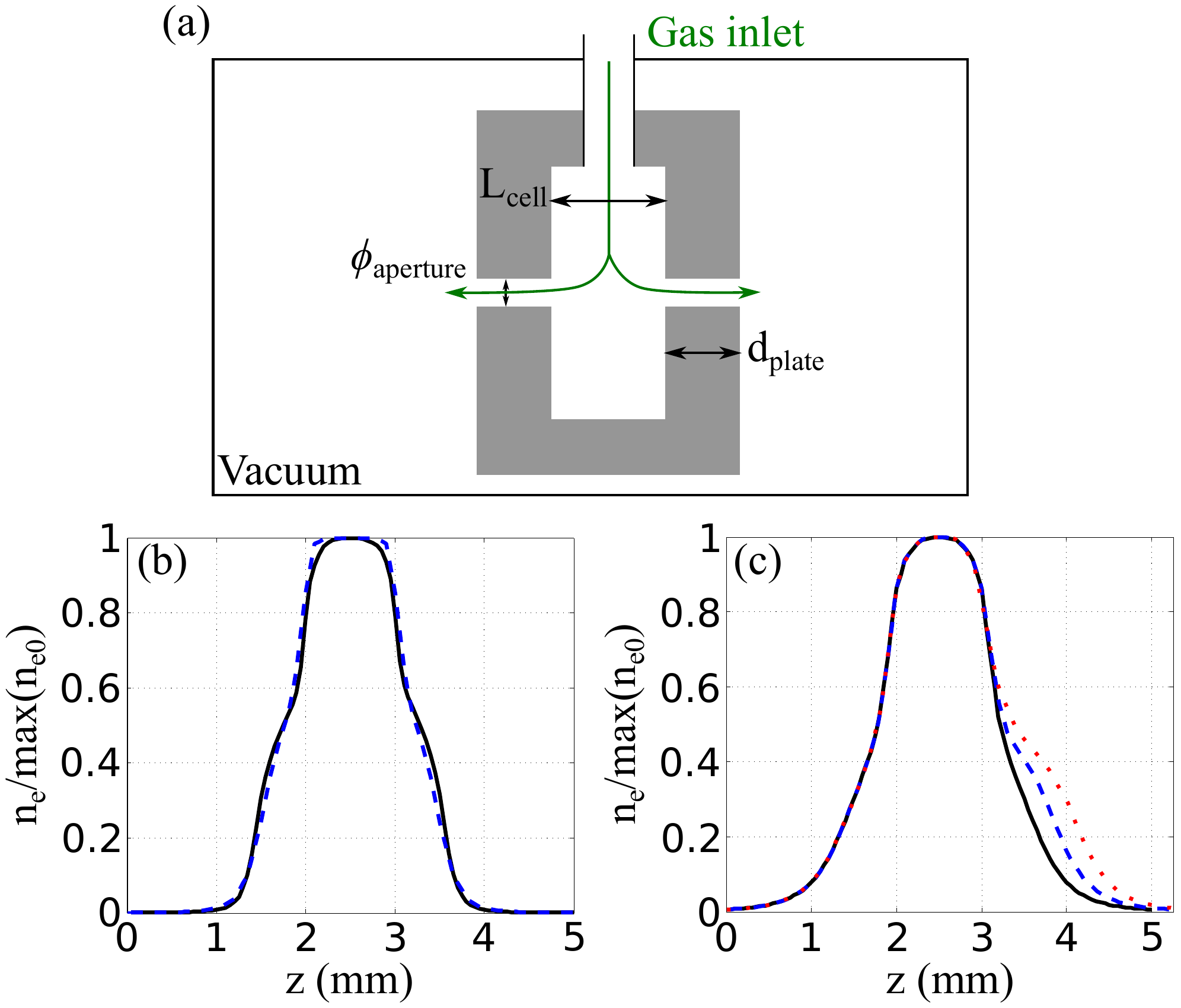}
\caption{(a) Schematic side view of gas cell. Normalized density profiles (b)  obtained using simulations with full 3D geometry (black solid line) and reduced geometry (blue dashed line) for  $L_{cell}=1\,\mathrm{ mm}$ and for plate of $200\, \mathrm{\mu m}$ aperture diameter and $500\, \mathrm{\mu m}$ thickness. (c) obtained with reduced geometry and $600\, \mathrm{\mu m}$ aperture diameter, $500\, \mathrm{\mu m}$ thick entry plate and $500\, \mathrm{\mu m}$ (black solid line) $750\, \mathrm{\mu m}$ (blue dashed line) and $1000\, \mathrm{\mu m}$ (red dotted line) thick exit plate. \label{fig:Fig1}}
\end{figure}

In our first calculations, the plasma target has a longitudinal density profile as represented by the black solid curve in Fig.~\ref{fig:Fig1}(c) with a $600\,\mathrm{\mu m}$ aperture diameter and $500\,\mathrm{\mu m}$ thickness plates at both entry and exit, as shown schematically in Fig.~\ref{fig:Fig1}(a).
This profile results from modeling the gas distribution inside an in-house gas cell used to conduct LWFA experiments, so-called ELISA \cite{audet_electron_2016}, which stands for ELectron Injector for compact Staged high energy Accelerator. The longitudinal plasma profiles of ELISA were characterized experimentally and by using dynamic fluid simulations with the SonicFoam solver in openFOAM \cite{weller_tensorial_1998}.

During LWFA experiments, laser ablation was observed to widen the diameter of the entry and exit plate apertures of the gas cell, resulting in an increase from $200\, \mathrm{\mu m}$ to $\approx 600\, \mathrm{\mu m}$. Hence $600\,\mathrm{\mu m}$ diameter apertures are chosen as initial values in order to ensure  conditions for gas flow independent of laser ablation. Simulations of gas flow with $600\,\mathrm{\mu m}$ diameter apertures were performed for several thicknesses of the exit plate aperture.

In order to reduce computational time and perform parametric studies, a reduced geometry of the cell was used for calculation of the longitudinal density profile. In this geometry, the outer cell diameter has a $2.5\,\mathrm{mm}$ radius around the laser axis instead of 20~mm. By keeping the cell length, $L_{cell}$, and the area of both the gas inlet and the apertures the same as the full geometry, a similar density profile is obtained as shown in Fig.~\ref{fig:Fig1}(b) for $200\, \mathrm{\mu m}$.

Fig.~\ref{fig:Fig1}(c) shows density profiles from simulations with $600\,\mathrm{\mu m}$ diameter plate aperture and different exit plate thickness. Comparison of Fig.~\ref{fig:Fig1}(b) and (c) shows that the up ramp and down ramp gradients using plate apertures of $600\,\mathrm{\mu m}$ diameter are smoother and longer as compared to plate apertures of $200\, \mathrm{\mu m}$ for the same thickness. Increasing the thickness of the exit plate aperture produces an extended, lower gradient at half of the maximum density, as shown in Fig.~\ref{fig:Fig1}(c) by the blue dashed line for a thickness of $750\,\mathrm{\mu m}$, and the red dotted line for $1000\,\mathrm{\mu m} $ thickness.

\subsection{Electron density determination}

The gas medium is composed of a mixture of hydrogen ($\mathrm{H_2}$) and nitrogen ($\mathrm{N_2}$). PIC simulations were performed for nitrogen concentration, $C_\mathrm{{N_2}}$ ranging from $0.35\%,\, 0.5\%,\, 1\%,\, 2\%$.

Due to the large difference in the ionization potential between the $5^{th}$ (L-shell) electron (IP $98\,\mathrm{eV}$) and the two K-shell electrons (IP 552 and $\mathrm{667\, eV}$) of the nitrogen atom, L-shell electrons will be ionized by the leading edge of the pulse where the intensity of the laser pulse is still rather moderate, whereas the ionization from the K-shell electrons occurs at higher intensities typically for $I>10^{18}\,\mathrm{W.cm^{-2}}$, so these electrons are born at rest in regions of strong fields, often at the laser peak intensity. As a consequence, at the intensities considered here, hydrogen and L-shell electrons of nitrogen contribute to the generation of the plasma wake and to the relativistic self-focusing of the laser while the K-shell electrons are ionized at a favorable wake phase and immediately trapped in the wakefields.

In order to study the influence of the amount of beam loading of  trapped electrons, it is essential to keep constant  the total number of electrons contributing to the plasma wake and to laser self-focusing. The total density of outer-shell electrons,  $n_{e}\left(z\right)$, is related to the density of atoms, $n_{at}$, by
\begin{align}
n_{e}\left(z\right) & =n_{at}\left(z\right)\left[\left(1-C_\mathrm{N_2}\right)+5C_\mathrm{N_2}\right]\nonumber \\
\hphantom{n_{e}\left(z\right)} & =n_{at}\left(z\right)\left[1+4C_\mathrm{{N_2}}\right].\label{eq:nenat}
\end{align}
From this equation, we observe that $n_{e}$ is increased by $8\%$ from the initial $n_{at}$ for $C_\mathrm{{N_2}}=2\%$. A variation in the target pressure of such percentage level can have a significant impact on laser-plasma interaction. In our PIC simulations, we assume that the target pressure is adjusted following Eq.~\ref{eq:nenat} so that the plasma density profile $n_{e}\left(z\right)$ becomes independent of $C_\mathrm{{N_2}}$.

\subsection{PIC simulations set-up}

For the results reported in this article, simulations were performed with WARP \cite{vay_novel_2012} using the azimuthal Fourier decomposition algorithm \cite{lifschitz_particle--cell_2009,davidson_implementation_2015,lee_modeling_2015} in cylindrical geometry. A field ionization module \cite{desforges_dynamics_2014} based on the ADK model \cite{ammosov_tunnel_1986} was introduced in WARP to model ionization dynamics.

A summary of the parameters used in our calculations is given in Table~\ref{tab:parameters}. The value of $a_{0}(z_f)=1.6$ corresponds to the maximum value of laser amplitude at the focal plane longitudinal position in vacuum, $z=z_{f}$.

\begin{table}[!htb]
\centering
\caption{\label{tab:parameters} List of parameters.}
\begin{tabular}{@{}lll@{}}
\hline
\hline
\begin{tabular}[c]{@{}l@{}}Maximum electron \\ number density on axis\end{tabular} & \multicolumn{1}{c}{$\max(n_{e0})$} & \multicolumn{1}{c}{$4\times 10^{18}\,\mathrm{cm^{-3}}$}  \\
Longitudinal density profile & & \multicolumn{1}{c}{ELISA profile}  \\
Plasma length  & \multicolumn{1}{c}{$L_{p}$} & \multicolumn{1}{c}{$5.0\,\mathrm{mm} $}   \\
   &    &   \\
Laser profile  & 						 & \multicolumn{1}{c}{$\mathrm{bi-Gaussian^{a}}$}  \\
Normalized vector potential & \multicolumn{1}{c}{$a_{0}(z_{f})$} & \multicolumn{1}{c}{$1.6$}  \\
Laser wavelength & \multicolumn{1}{c}{$\lambda_0$} & \multicolumn{1}{c}{$0.8\,\mathrm{\mu m}$}   \\
Laser waist  & \multicolumn{1}{c}{$w^{b}$} & \multicolumn{1}{c}{$16\,\mathrm{\mu m}$}  \\
Laser duration (FWHM) & \multicolumn{1}{c}{$\tau$}  & \multicolumn{1}{c}{$20\,\mathrm{fs}$}  \\
Laser focal position & \multicolumn{1}{c}{$z_{f}$}  & \multicolumn{1}{c}{$2.8\,\mathrm{mm}$}   \\
Laser polarization  &    & \begin{tabular}[c]{@{}c@{}}linear\\ (in $y-$direction)\end{tabular} \\
   &    &   \\
Number of Fourier modes &    & \multicolumn{1}{c}{$2$}   \\
Number of particles/cell/species &    & \multicolumn{1}{c}{$36\, \mathrm{macro}$}  \\
\hline
\hline
\end{tabular}
\begin{flushleft}
$^{a}$Gaussian in temporal and spatial profiles
\\
$^{b}$Radius of the beam at $1/e^2$ of the laser intensity
\end{flushleft}
\end{table}

The mesh resolution was chosen to be $\Delta z =\lambda_0/25$ and $\Delta r= \lambda_0/6$ in the longitudinal and transverse directions. A mesh resolution of $\Delta z =\lambda_0/30$ and $\Delta r= \lambda_0/10$ was used for the most optimum case for confirmation and for a more precise evaluation on the second-order beam properties, such as the beam emittance.

\section{Influence of nitrogen concentration on trapped electron dynamics}
\label{sec:analysis}
Simulations with parameters previously shown in Sec.~\ref{sec:numerical} were performed. In this section we discuss the role of nitrogen concentration on the properties of the resulting electron beam.

\subsection{Laser plasma interaction}

\begin{figure}[htb]
\includegraphics{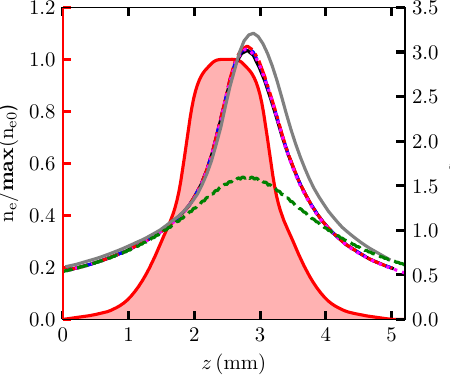}

\caption{Electron density profile along laser propagation axis (left vertical axis, red curve and shaded area), identical to the black solid curve in Fig.~\ref{fig:Fig1}(c). Evolution of $a_{0}(z)$ along propagation axis $z$ (right vertical axis)  in vacuum (green dashed line) and in plasma at different $C_\mathrm{N_2}$: $2\%$ (black solid line), 1\% (blue dashed line), 0.5 \% (red dashed-dotted line) and 0.35 \% (magenta dotted line). The gray solid line represents the evolution of $a_{0}(z)$ with respect to the propagation axis $z$ for slightly different laser parameters,  beam waist $17\,\mathrm{\mu m}$ and  focal position at $2.9\,\mathrm{mm}$, discussed in Section~\ref{sec:laser}. \label{fig:fig2}}
\end{figure}

The evolution of $a_0(z)$ (right vertical axis) as a function of $z$ is plotted in Fig.~\ref{fig:fig2} during propagation in a plasma with the ELISA longitudinal density profile (left vertical axis, red curve and colored area, same as the black solid curve in Fig.~\ref{fig:Fig1}(c); this profile is kept identical when varying $C_\mathrm{N_2}$. In vacuum (green dashed curve in Fig.~\ref{fig:fig2}), $a_0$ reaches $1.6$ at the focus position,$z_f=2.8\,\mathrm{mm}$;  in the plasma,  self-focusing increases the value of $\max(a_{0}(z))$ by nearly a factor 2 compared to the vacuum case.

Fig.~\ref{fig:fig2} shows that the evolution of  $a_0(z)$ is the same for   $ C_\mathrm{N_2}\leq 2\%$, implying that plasma wave generation is the same for these cases where $n_{e0}=\max\left[n_{e}\left(z\right)\right]$ is kept constant, and the influence of the concentration on electron dynamics can be studied independently of other parameters. The two K-shell electrons of nitrogen are the only contributors to the trapped charge in the plasma wave, this implies that $C_\mathrm{{N_2}}$ is a key parameter for controlling the amount of beam loading during trapping and acceleration processes.

Ionization and trapping of electrons coming from the ionization process $\mathrm{N}^{5+}\rightarrow \mathrm{N}^{6+}$ take place efficiently for $a_{0}>1.5$, starting at $z=2.1\, \mathrm{mm}$, in the density up ramp, close to the density plateau. The trigger of $\mathrm{N}^{6+}\rightarrow \mathrm{N}^{7+}$ ionization process requires an $a_{0}\geq2.0$, occurring at $z=2.3\,\mathrm{mm}$, $200\,\mathrm{\mu m}$ further than in the previous case. As we will see later, this $200\,\mathrm{\mu m}$ difference in the initial position of trapping has a strong influence on the properties of accelerated electrons.

In the following, we denote  $6^{th}$ (respectively $7^{th}$) electron, the electron created from the ionization process $\mathrm{N^{5+}}\rightarrow\mathrm{N^{6+}}$ (resp. $\mathrm{N^{6+}}\rightarrow\mathrm{N^{7+}}$).\\

Simulations were performed up to a few hundreds of $\mu m$ after the exit plate of the plasma target ($z= 3.5\,\mathrm{mm}$), at positions where the plasma wakefield is nearly zero. At these positions, the divergence of the electron beam has already reduced the space charge force, therefore electrons propagate nearly without interaction.

\subsection{Electron dynamics during acceleration}

\subsubsection{Analysis of ionization and trapping mechanisms}

\begin{figure}[htb]
\includegraphics{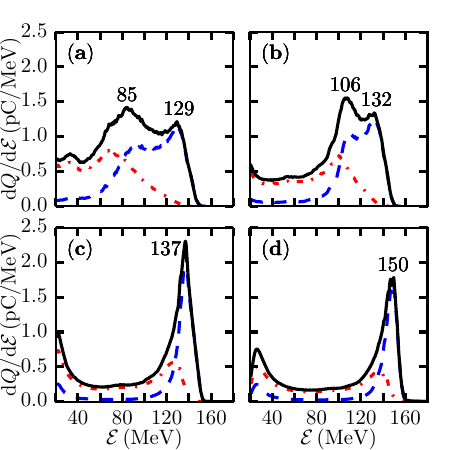}
\caption{Energy distribution of electrons having an energy $\geq20\, \mathrm{MeV}$ at the exit of the plasma target. Each figure corresponds to a specific value of $C_\mathrm{N_2}$, representing the proportion of nitrogen
in the gas cell. (a) $C_\mathrm{N_2}=2\%$; (b) $C_\mathrm{N_2}=1\%$;
(c) $C_\mathrm{N_2}=0.5\%$; (d) $C_\mathrm{N_2}=0.35\%$. In these
figures, the blue dashed line shows the contribution of electrons coming
from the ionization process $\mathrm{N}^{5+}\rightarrow\mathrm{N^{6+}}$, while the red dashed-dotted line is for the ionization process $\mathrm{N}^{6+}\rightarrow\mathrm{N^{7+}}$. The sum of these two contributions is represented by the black solid line. Energy peak values are written on top of energy peaks.\label{fig:Fig3} }
\end{figure}

The electron distribution is plotted as a function of their energy in Figs.~\ref{fig:Fig3}(a-d) for the four values of $C_\mathrm{N_2}$; the dashed-blue (dash-dotted-red) curves represent the contribution of the $6^{th}$ (respectively, $7^{th}$) electrons, and the black solid curve represents the sum of all electrons. For the laser-plasma configuration  described in Tab.~\ref{tab:parameters}, there is no electron coming from either hydrogen or the L-shell of nitrogen among those accelerated to high energies. As can be seen from Figs.~\ref{fig:Fig3}(a-d), the highest beam energy range is close to $150\,\mathrm{MeV}$. However, the obtained energy distributions change from broad spectra to peaked distributions with pedestal for $C_\mathrm{N_2} < 1\%$, indicating that $C_\mathrm{N_2}$ has a strong influence on the dynamics of electrons in the trapping and acceleration processes.

In Fig.~\ref{fig:Fig3}, the total energy distribution exhibits two  peaks in (a) and (b) and single peaks  in (c) and (d). The high energy peaks, at $129\,\mathrm{MeV}$ and $132\,\mathrm{MeV}$ for cases (a) and (b) respectively, are close to the ones of cases (c) and (d).  As seen from the blue-dashed curves, the $6^{th}$ electrons contribute dominantly to the high energy range, whereas the $7^{th}$ electrons constitute the tail of the bunch because they are generated at higher laser intensity, so that their trapping occurs at a later time during laser-plasma interaction. The charge density of the $6^{th}$ electrons, $\mathrm{d}Q/\mathrm{d}\mathcal{E}$, decreases to almost zero for $\mathcal{E}$ less than the energy of the $7^{th}$ electron peak; on the contrary, the charge density of the $7^{th}$ electrons exhibits a pedestal, with a value comparable to its maximum. These structures result from the history of ionization over the plasma length. In fact, as soon as the laser vector potential is high enough to give rise to the ionization process $\mathrm{N^{5+}}\rightarrow\mathrm{N^{6+}}$\textbf, a significant amount of the $6^{th}$ electrons is trapped in the first bucket of the plasma wave, resulting in a reduction of the trapping potential at the back of the bucket due to beam loading effects. Simultaneously, the laser self-focuses gradually while interacting with the plasma, $a_0$ increases and exceeds the intensity threshold for the ionization process $\mathrm{N^{5+}}\rightarrow\mathrm{N^{6+}}$. This leads to the laser pulse head becoming intense enough to generate the $6^{th}$ electrons, at positions where the relative value of the plasma wake potential is still rather high, or the trapping probability is low. As a consequence, these electrons either cannot be trapped, or are trapped at the back of the bucket. The $7^{th}$ electrons are generated close to the maximum laser intensity, and are thus more easily trapped, and continuously injected into the plasma wakefield, creating a broad energy distribution.

\subsubsection{Analysis of beam loading effects }

\begin{figure}[ht]
\includegraphics{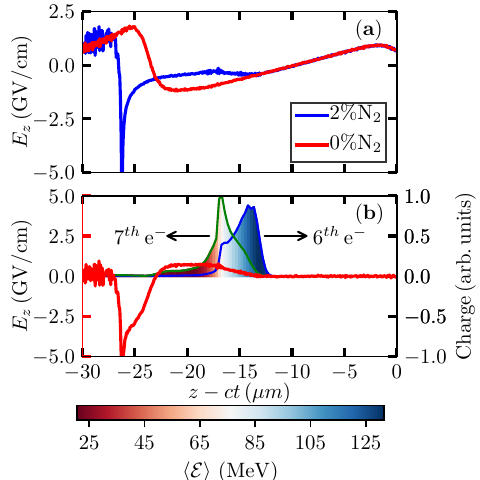}
\caption{Illustration of beam loading effects taken at $z=3.3\,\mathrm{mm}$. (a)  $E_z$ fields with $2\% $ of $\mathrm{N_2}$ in blue solid curve and with $0\% $ of $\mathrm{N_2}$ in red solid curve with respect to $z-ct$; (b)  Beam loaded electric field in red solid curve given by subtracting $E_z$ fields with $0\% $ of  $\mathrm{N_2}$ from the one with $2\%$ of $\mathrm{N_2}$. In blue solid line is the charge distribution of the $6^{th}$ electrons and in green solid line is the charge distribution of the $7^{th}$ electrons. The color bar represents the average energy distribution of electrons, with a bin size along $z-ct$ of $ 0.7\,\mathrm{\mu m}$. High energy electrons are located at the front of the bunch. \label{fig:fig18}}
\end{figure}

Fig.~\ref{fig:fig18} illustrates the importance of beam loading effects, induced by trapped electrons, which number varies with $C_\mathrm{N_2}$ concentration. Fig.~\ref{fig:fig18}(a) shows the longitudinal electric field $E_z$ for cases with $C_\mathrm{N_2}=2\%$ in blue solid line and without nitrogen atom in red solid line, and the red solid curve in Fig.~\ref{fig:fig18}(b) shows the induced field by trapped electrons (left vertical axis) determined by subtracting $E_z$ without nitrogen from the one with $C_\mathrm{N_2}=2\%$. Charge distributions (right vertical axis) along $z-ct$ for $6^{th}$ and $7^{th}$ electrons are given in blue and green solid lines, respectively. The color bar represents the average energy of electrons in each slice along $z-ct$.

The amount of beam loading, generated by the high energy electron bunch gradually increases from the front of the bunch ($z-ct=-12\,\mathrm{\mu m}$) up to its back ($z-ct=-23\,\mathrm{\mu m}$). Electrons at the front of the bunch experience an amount of beam loading close to zero, hence their maximum energy is preserved.
The  energy distribution along $z$   in Figs.~\ref{fig:Fig3}(b) shows that the $6^{th}$ electrons are trapped at the earliest time of the laser-plasma interaction process, and that they remain at the front of the bunch up to the exit of the plasma target, whereas the $7^{th}$ constitute the tail of the bunch with a lower energy. As soon as the electrons get enough energy (i.e. $>20\, \mathrm{MeV}$), their relative position remains invariant during the acceleration process.

The  peak at $85 \,\mathrm{MeV}$ for $C_\mathrm{N_2}=2\%$ moves to $106\,\mathrm{MeV}$ for $C_\mathrm{N_2}=1\%$ as shown in Figs.~\ref{fig:Fig3}(a-b), and finally at $C_\mathrm{N_2}=0.5\%$, merges with the high energy peak at $137\,\mathrm{MeV}$, as shown in Fig.~\ref{fig:Fig3}(c); this process indicates a correlation between the formation of the peaks and $C_\mathrm{N_2}$, through beam loading effects. The beam-loaded electric field reduces the amplitude of the net accelerating field, which is flattened. As a result, as shown by the  distribution of electrons in energy and position plotted in Fig.~\ref{fig:fig18}(b), lower energy electrons situated at the tail of the bunch,  never catch up with the higher energy electrons situated at the head, resulting in a large separation between low and high energy peaks, leading to a broad energy spectrum.

When $C_\mathrm{N_2}<0.5\%$, the amount of beam loading is less significant, the accelerating gradient along the accelerated bunch is larger, therefore electrons at the tail of the bunch eventually gain more energy than electrons at the head. This phenomenon begins to be visible for $C_\mathrm{N_2}=0.35\%$, in particular through the presence of two neighboring peaks at the maximum of the black curve in Fig.~\ref{fig:Fig3}(d). Reducing $C_\mathrm{N_2}$ to $<0.35\%$ will thus lead to an increase of the energy spread in the high energy peak. In summary, for the considered parameters, the energy spread of the electron beam is minimized for a value of $C_\mathrm{N_2}$ between $0.5$ and $0.35\%$.

\subsubsection{Influence of $C_\mathrm{N_2}$ on the evolution of energy spread}

\begin{figure}[htb]
\includegraphics{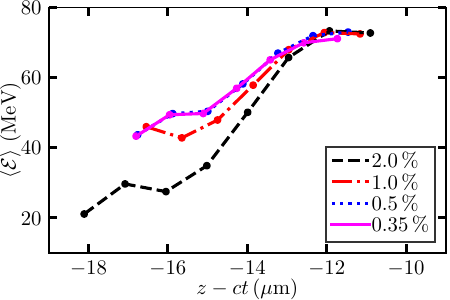}

\caption{Average energy $\left<\mathcal{E}\right>$ of electrons situated in the FWHM of the energy distribution of accelerated electrons at the early phase of acceleration, around $ z= 2.7\,\mathrm{mm}$. \label{fig:Fig.4}}
\end{figure}

\begin{figure}[htb]
\includegraphics{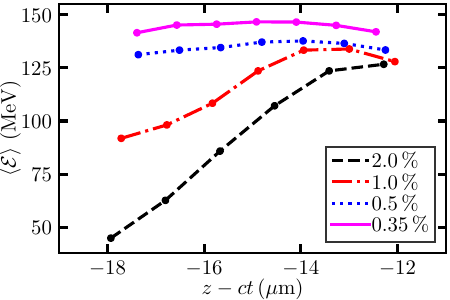}
\caption{Same as Fig.~\ref{fig:Fig.5} at the exit of the gas cell.
\label{fig:Fig.5}}
\end{figure}

The parameter $C_\mathrm{N_2}$ is  efficient for controlling  phase space rotation, which is essential for the control of  energy spread. The average energy $\left\langle \mathcal{E}\right\rangle$ of high energy electrons is plotted as a function of $z-ct$ for the four values of $C_\mathrm{N_2}$, at the early phase of bunch acceleration, around $2.7\,\mathrm{mm}$ in Fig.~\ref{fig:Fig.4}, and at the plasma exit in Fig.~\ref{fig:Fig.5}. At  $z \sim 2.7\,\mathrm{mm}$ the energy $\left\langle \mathcal{E}\right\rangle $ increases monotonically with $z-ct$. Electrons trapped earlier are located at larger values of $z-ct$, and have been accelerated over longer distance, gaining more energy. Regardless of $C_\mathrm{N_2}$, the ratio of energy  between the head and the tail of the bunch is $>2$.

At the exit of the plasma, the ratio between the maximum and the minimum values of the curves decreases with $C_\mathrm{N_2}$ from 3.12 to 1.06. This shows that the decrease of $C_\mathrm{N_2}$ induces an increase of phase space rotation, minimizing  energy spread. The phase space distribution is more symmetrical for $C_\mathrm{N_2}\leq0.5\%$ as compared to the one of $C_\mathrm{N_2}\geq1\%$, implying that for low $C_\mathrm{N_2}$ electrons at the head and tail attain similar average energy $\left<\mathcal{E}\right>$.  For $C_\mathrm{N_2}=0.35\%$, the distribution for $z-ct < -14\,\mathrm{\mu m}$ is no longer increasing monotonously, implying that the degree of phase space rotation begins to be too large. A further decrease of $C_\mathrm{N_2}$ will yield an additional degree of phase space rotation, degrading the energy spread.

It shows, in agreement with previous results \cite{lee_dynamics_2016}, that for a given laser-plasma configuration, there is a specific value of the percentage of trace atoms minimizing the energy spread. In \cite{lee_dynamics_2016}, this value was $C_\mathrm{N_2}=1\%$, whereas in the present study it occurs at $C_\mathrm{N_2}\simeq0.5\%$. This phenomenon is also observed in experiments \cite{couperus_demonstration_2017}. Note that $\max(n_{e0})$ is also reduced by a factor of two as compared to the one in \cite{lee_dynamics_2016}. Hence,  the accelerating field gradient suitable to achieve the smallest dependency between  energy and trapping position can be obtained by decreasing $C_\mathrm{N_2}$ with the plasma density.

\subsection{General beam properties at the exit of the plasma target}

In this section we summarize the properties of the high energy electron bunch generated by the laser plasma injector with $C_\mathrm{N_2}$ as a parameter. We consider  a bunch of electrons with energy centered at the value of the high energy peak $\mathcal{E}_{peak}$ as reported in Figs.~\ref{fig:Fig3}(a-d) and distributed  over the full width at half maximum (FWHM) of the peak.

\subsubsection{Beam energy}
\begin{figure}[htb]
\includegraphics{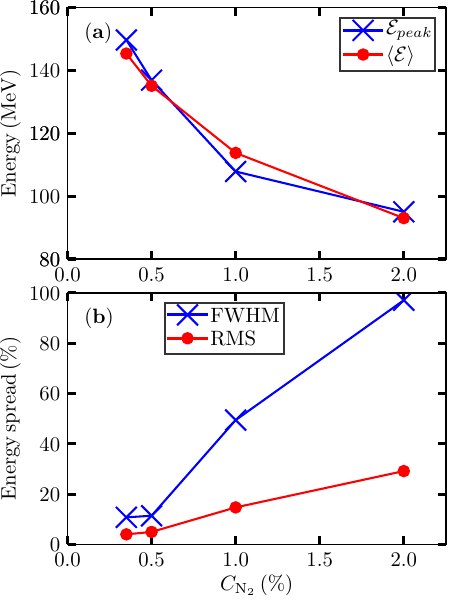}
\caption{Accelerated beam energy properties for each $C_\mathrm{N_2}$. (a) Peak energy $\mathcal{E}_{peak}$ (blue crosses) and average energy $\left<\mathcal{E}\right>$ (red dots) with respect to $C_\mathrm{N_2}$; (b) Energy spread in full width at half maximum (FWHM), represented by blue crosses and root-mean-square (RMS), represented by red dots with respect to $C_\mathrm{N_2}$.\label{fig:Fig7}}
\end{figure}

In Fig.~\ref{fig:Fig7}(a) the peak energy, $\mathcal{E}_{peak}$, and the average energy, $\left<E\right>$, are plotted as functions of $C_\mathrm{N_2}$. The decrease of $\mathcal{E}$ with increasing $C_\mathrm{N_2}$ is explained by the fact that the beam-loaded electric field reduces the accelerating field of the plasma wave, resulting in a decrease of energy gain. The beam-loaded electric field has a significant impact at  $C_\mathrm{N_2}$ as low as $1\%$, as a drastic diminution of $\mathcal{E}_{peak}$ is observed between $0.5\%$ and $1\%$. For $C_\mathrm{N_2} = 0.5 \%$ and $0.35 \%$, the average energy $\left<\mathcal{E}\right>$, $135$ and $145\, \mathrm{MeV}$ respectively, is in the specified energy range.

Fig.~\ref{fig:Fig7}(b) shows the energy spread of the electron bunch comprised within the FWHM of the energy spectrum, with $\Delta \mathcal{E}_\mathrm{FWHM}/\left<\mathcal{E}\right>$ the energy spread evaluated at FWHM and $\Delta \mathcal{E}_{rms}/\left<\mathcal{E}\right>$, evaluated at root-mean-square. Both quantities increase dramatically when $C_\mathrm{N_2}\geq1\%$, with $\Delta \mathcal{E}_\mathrm{FWHM}/\left<E\right>=50\%$ at $C_\mathrm{N_2}=$1\% and close to $100\%$ at $C_\mathrm{N_2}=2\%$. At a lower concentration, $\Delta \mathcal{E}_\mathrm{FWHM}/\left<\mathcal{E}\right>$ is close to $10\%$ for $C_\mathrm{N_2}=0.35-0.5\%$, while $\Delta \mathcal{E}_{rms}/\left<\mathcal{E}\right>$ are 4\% at $C_\mathrm{N_2}=0.35\%$ and 5\% at $C_\mathrm{N_2}=0.5\%$.

\subsubsection{Beam charge}
\begin{figure}[htb]
\includegraphics{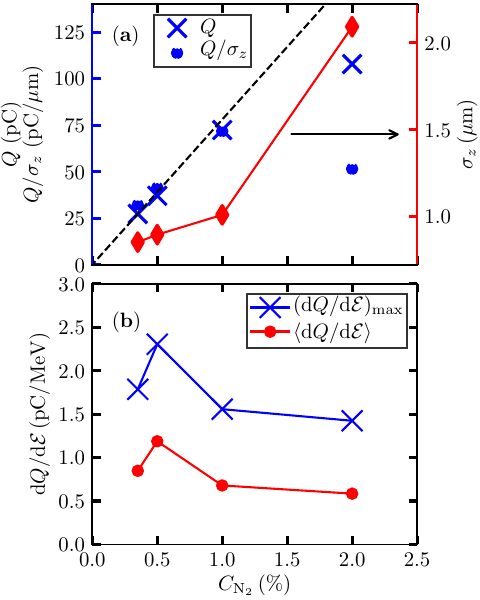}

\caption{Accelerated beam charge with respect to $C_\mathrm{N_2}$. (a) On the left axis: Charge $Q$ (blue crosses) and linear charge density $Q/\sigma_z$ (blue dots) on the right axis with respect to $C_\mathrm{N_2}$, the black dashed lie represents a linear regression of $Q$ with respect to $C_\mathrm{N_2}$ passing by the value
$Q$ at $C_\mathrm{N_2}=0.35\%$. On the right axis: Electron rms bunch length $\sigma z$ with respect to  $C_\mathrm{N_2}$; (b) Maximum charge density $(\mathrm{d}Q/\mathrm{d}\mathcal{E})_{max}$ (blue crosses) and average charge density $\left<\mathrm{d}Q/\mathrm{d}\mathcal{E}\right>$ (red dots). \label{fig:Fig.9}}
\end{figure}

In Fig.~\ref{fig:Fig.9}(a), the absolute value of the charge, $Q$, is plotted with blue crosses, the ratio of absolute value of the charge to the rms electron bunch length $Q/\sigma_z$ as blue dots, and on the right-hand axis, the rms electron bunch length $\sigma_z$  as red diamonds as a function of $C_\mathrm{N_2}$. As expected the charge increases with $C_\mathrm{N_2}$. The comparison with a linear regression, given by the black dashed-line, shows however a saturation effect, which becomes visible at $C_\mathrm{N_2}=1\%$ and significant at $C_\mathrm{N_2}=2\%$. For $C_\mathrm{N_2}=0.5$\% and 1\%, we obtain $Q=27.3$ and $37.0\mathrm{\, pC}$ respectively, these values meet the charge requirement for application of optical injector in a multistage accelerator.

The length of the electron bunch has a direct influence on the beam-loaded electric field.
The rms bunch duration   corresponding to Fig.~\ref{fig:Fig.9}(a) in time units is given by $\sigma_{t}(fs) = 3.34 \sigma_{z}(\mathrm{\mu m})$. For $C_\mathrm{N_2}$= 0.35 and 0.5 \%, $\sigma_{z}= 0.85$ and $0.89\, \mathrm{\mu m}$ respectively, the corresponding $\sigma_{t}$ are $2.8$ and $3.0\, \mathrm{fs}$ respectively. These durations are only 5\% of the minimum plasma period $\tau_{p}=56 fs$, reached for $n_{e}=4\times10^{18}\,$cm$^{-3}$. These extremely short electron bunches are well adapted for injection in an accelerating stage for further acceleration.  Fig.~\ref{fig:Fig.9}(a), shows  that the duration increases quasi-linearly between $0.35$ and $1\%$ of nitrogen concentration, then increases drastically when  $C_\mathrm{N_2} = 2\%$. The increase of the bunch length with $C_\mathrm{N_2}$ contributes directly to the increase of the energy spread as observed in Figs.~\ref{fig:Fig3} and \ref{fig:Fig7}(b). In Figs.~\ref{fig:Fig.9}(a), the blue circles represent the value of $Q/\sigma_{z},$ which evolution is equivalent to the linear density of beam charge $Q$. We observe a maximum at $C_\mathrm{N_2}=1\%$, indicating that above a given threshold, here around $70\,\mathrm{ pC}$, an increase of the trapped charge $Q$ results in an increase of its length $\sigma_z$, and a decrease of its linear charge density $Q/\sigma_z$.

In Fig.~\ref{fig:Fig.9}(b) are reported the maximum value of the charge density per energy unit $\left(dQ/dE\right)_{max}$ and its average defined by $\left<{dQ/dE}\right>=Q/(4\Delta \mathcal{E}_{rms})$. Both quantities have the same trends, the highest values are obtained for $C_\mathrm{N_2}=0.35$ and $0.5\%$ with the corresponding values $\left(dQ/dE\right)_{max}= 1.8$ and $2.3 \,\mathrm{pC/MeV}$ and $\left<{dQ/dE}\right>$ = 1.2 and 1.4 $\mathrm{pC/MeV}$.

\subsubsection{Beam divergence and emittance}

\begin{figure}
\includegraphics{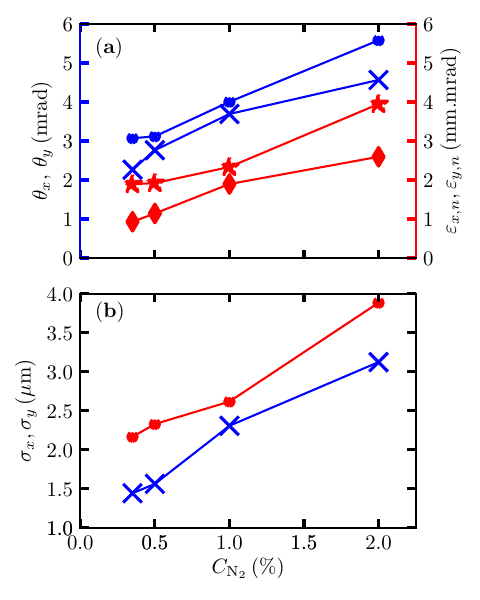}

\caption{Accelerated beam divergence and emittance with respect to $C_\mathrm{N_2}$. (a) On the left axis: rms divergence $\theta_x$ in $x$-direction (blue crosses), and $\theta_y$ in $y$-direction (blue dots) with respect to $C_\mathrm{N_2}$. On the right axis: normalized rms emittance $\varepsilon_{x,n}$ in $x$-direction (red diamonds), and $\varepsilon_{y,n}$ in $y$-direction (red stars) with respect to $C_\mathrm{N_2}$; (b) Apparent electron bunch length $\sigma_x$ in $x$-direction (blue crosses), and $\sigma_y$ in $y$-direction (red dots) with respect to $C_\mathrm{N_2}$. \label{fig:Fig.11}}

\end{figure}

The beam divergence and emittance in the transverse plane ($x,y$), $y$ being the laser polarization axis, are plotted as a function of $C_\mathrm{N_2}$ in Fig.~\ref{fig:Fig.11}(a).

The normalized rms emittance along $i$-axis, $\varepsilon_{i,n}$ is calculated using the standard formula:
\begin{equation}
\varepsilon_{i,n}^2=\left\langle x_i^{2}\right\rangle \left\langle \frac{p_{i}}{m_ec}\right\rangle ^{2}-\left\langle x_i\frac{p_{i}}{m_ec}\right\rangle ^{2},
\end{equation}
where $x_i$ and $p_{i}$ are the electron position and momentum along the $i$-axis.  As can be seen in Fig.~\ref{fig:Fig.11}(a), both rms angular divergence and normalized rms emittance increase with $C_\mathrm{N_2}$, the ratio between their maximum (at $C_\mathrm{N_2}=2\%$) and minimum (at $C_\mathrm{N_2}=0.35\%$) being close to a factor of 2. The difference between  $x-$ and $y-$directions is specific to the ionization induced injection process. After  ionization, the electrons gain momentum (positive or negative) in the laser polarization direction. This gain is responsible for the increase of emittance along the laser polarization axis. Nevertheless, the values obtained for $C_\mathrm{N_2}=0.35$ and $0.5 \%$ appear to be acceptable for an injector of a multistage accelerator, the corresponding values are $\theta_{x}= 2.3$ and $2.8\,\mathrm{mrad}$, $\theta_{y}= 3.0$ and $3.1 \,\mathrm{mrad}$, $\varepsilon_{x,n}= 0.93$ and $1.14\,\mathrm{mm.mrad}$, $\varepsilon_{y,n}=1.89$ and $1.92\, \mathrm{mm.mrad}$ respectively. These values of $\varepsilon_{y,n}$ are in accordance with the measured ones in \cite{barber_measured_2017}. In \cite{barber_measured_2017}, with a charge density of $\sim2\,\mathrm{pC/MeV}$ as in our simulation, the measured emittance is $\sim 1.8\,\mathrm{mm.mrad}$ in the laser polarization direction, contributed by the intrinsic source emittance and the beam loading effects. From the obtained results, the main limiting factor for the performance of an injector appears to be the value of $\varepsilon_{y,n},$ close to $2\,\mathrm{mm.mrad}$. This issue can be handled by designing a specific magnetic transport line to reduce the emittance before injection to the accelerating stage.

The rms apparent transverse size of the accelerated electron bunch can be estimated at large distances from the exit of the plasma. Due to the divergence of the beam, at large distances the beam-loaded electric field becomes negligible, thus electrons propagate freely in space. For such propagation, both the geometric emittance and the angular divergence of the electron beam are constant quantities. Using the emittance definition
\begin{equation}
\varepsilon_{x}=\sqrt{\left\langle x^{2}\right\rangle \left\langle \theta_{x}^{2}\right\rangle -\left\langle x\theta_{x}\right\rangle ^{2}},\label{eq:emittanceX}
\end{equation}
with $\theta_{x}$ =$v_{x}/v_{z}$, $v_{x}, v_{z}$ being the electron velocity on the $x,z$-axes, the rms apparent transverse size $\sigma_{x}$ of the source of electrons along the $x$-axis is determined as the minimum value of $\sqrt{\left\langle x^{2}\right\rangle }$. The previous shown formula is for the emittance in the $x-$direction, it is also valid in the $y-$direction. Assuming free propagation of electrons the minimum value of $\sqrt{\left\langle x^{2}\right\rangle }$ is obtained when $d\left\langle x^{2}\right\rangle /dt =0$, or $\left\langle x v_{x}\right\rangle =0.$ Electrons in the high energy peak are highly relativistic, we can therefore assume $v_{z}\sim c$ and $v_{x}=c\theta_{x}.$ As a consequence, the minimum value of $\sqrt{\left\langle x^{2}\right\rangle }$ is obtained for $\left\langle x\theta_{x}\right\rangle =0$. Substituting this relation in Eq.~\ref{eq:emittanceX}, we obtain
\begin{equation}
\sigma_{x} =\frac{\varepsilon_{x}}{\theta_{x}},\label{eq:SourceSize}
\end{equation}
where $\theta_{x}$ is the rms value of the angular divergence given in Fig.~\ref{fig:Fig.11}. An equation similar to Equation~\ref{eq:SourceSize} can also be used along the $y$-axis. $\sigma_{x}, \sigma_{y}$ are given in Fig.~\ref{fig:Fig.11}(b). The data show the same general trends as in Fig.~\ref{fig:Fig.11}(a): the rms apparent transverse size of the electron bunch increases with $C_\mathrm{N_2}$ and  larger values are achieved along the polarization axis. At $C_\mathrm{N_2}=0.35$ and $0.5\%$, the corresponding values are $\sigma_{x}=1.44$ and $1.56 \,\mathrm{\mu m}$, and $\sigma_{y}=2.17$ and $2.33 \,\mathrm{\mu m}$. We have checked that the beam-loaded electric field is negligible at the exit of the plasma target, and estimate that the emittance is close to 90\% of its asymptotic value for $C_\mathrm{N_2}= 2\%$.

\section{Influence of  density profile}
\label{sec:downramp}
As shown in Section~\ref{sec:analysis}, in the selected laser plasma configuration, optimum electron bunch  properties are optimized for $C_\mathrm{N_2}$= 0.35 and 0.5\%. In this section,  the sensitivity of these properties on the longitudinal density profile is analysed.

As ionization injection starts close to the plateau of the longitudinal density profile (see Fig.~\ref{fig:fig2}),  the particular shape of the density profile at the entrance of the plasma is not significant. A modification of this profile can be compensated by varying the focal plane position of the laser beam, in order to get nearly the same evolution at the beginning of the plateau. However, the longitudinal density profile at the exit of the plasma affects the dynamics of  generated electrons. In this section, we analyze the effect induced by a modification in the longitudinal density profile.

\subsection{Evolution of beam properties along the  plasma profile}
\label{subsec:evolution}

\begin{figure}[htb]
\includegraphics[width=8cm]{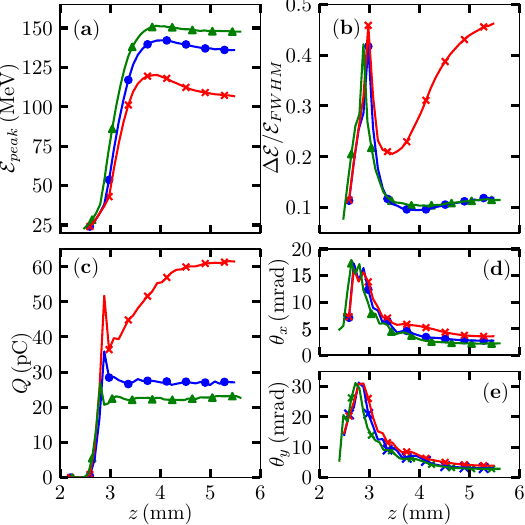}

\caption{Evolution of four beam quantities over the longitudinal traveling distance: (a) Peak energy ($\mathcal{E}_{peak}$); (b) FWHM energy spread ($\Delta\mathcal{E}/\mathcal{E}$); (c) Charge ($Q$) of electrons; (d) rms values of the angular divergence in $x-$ and (e) in $y-$ directions. In each figure, three curves are represented, each corresponds to a concentration value: $1\%$ (red crosses), $0.5\%$ (blue dots), and $0.35\%$ (green triangles).\label{fig:Fig.14}}

\end{figure}

Fig.~\ref{fig:Fig.14} shows the evolution  of  (a) $\mathcal{E}_{peak}$, (b) $\Delta \mathcal{E}_{FWHM}/\mathcal{E}_{peak}$, (c) $Q$, (d) $\theta_x$, and (e) $\theta_y$, during the acceleration phase, for concentration values of $1\%$ (red crosses), $0.5\%$ (blue dots), and $0.35\%$ (green triangles).  $z=2.6\, \mathrm{mm}$  corresponds to the initial position where the high energy peak can be well identified.

Fig.~\ref{fig:Fig.14}(a) shows that for all values  of $C_\mathrm{N_2}$, the maximum energy of the peak is obtained at about $z= 3.5\,\mathrm{ mm}$. As shown in Fig.~\ref{fig:fig2} at this position, the plasma density is reduced by a factor of two compared to the value at plateau.  Moreover a large increase of energy occurs between $z=2.8 \,\mathrm{mm}$, corresponding to the end of the plateau, and $z=3.5\, \mathrm{mm}$. This indicates that a large part of the acceleration occurs within the down ramp. As can be observed from Fig.~\ref{fig:Fig.14}(c), the variation of  beam charge for $C_\mathrm{N_2}=0.5$ and $0.35\%$ is minimal after $z= 3\,\mathrm{ mm}$: as soon as the high energy peak is formed, which occurs at energies much lower than the final one, the charge of the high energy peak is nearly constant, and injection and acceleration zones are thus well separated. This is the main reason why a low energy spread is achieved for these two concentration values, as shown in Fig.~\ref{fig:Fig.14}(b). Although  new electrons are  trapped in the acceleration zone, in particular the $7^{th}$ electrons, they are located behind the high energy peak position, where the accelerating field is moderate. Beyond the position of maximum energy, for $z> 3.5\,\mathrm{mm}$, the longitudinal and transverse fields can still be large enough to modify the characteristics of the high energy electron bunch. Although the high energy electrons are still in the accelerating zone of the plasma wakefield, there is no more global acceleration of the electron bunch after this maximum, signifying that the beam loading effects cancel out, or overcome, the plasma wakefield. This can be seen by comparing the maximum energy at $z=3.5\,\mathrm{mm}$ and the final one obtained  at $z=5.5\,\mathrm{mm}$ for each concentration value shown in Fig.~\ref{fig:Fig.14}(a). For $C_\mathrm{N_2} =1\%$, there is a reduction of $11\%$ after the maximum energy is reached; for $C_\mathrm{N_2}$=0.5 \%, a reduction of $4\%$; for $C_\mathrm{N_2}= 0.35 \%$, a reduction of $2\%$. In addition, the energy spread, shown in Fig.~\ref{fig:Fig.14}(b), is strongly correlated with the energy reduction. A large reduction of energy, dominated by beam-loaded electric field, results in a large energy spread. The transverse field at $z\geq 3.5\,\mathrm{mm}$,  yields a significant decrease of the angular divergence for all three values of $C_\mathrm{N_2}$.

\subsection{Variation of down ramp profile}
\label{sec:downramp}
Results of Section \ref{subsec:evolution} demonstrate that the down ramp of the longitudinal density profile contributes significantly to several beam properties. Here we analyze the influence of its variation by extending the plate  thickness at the exit from $500$ to $750$ and $1000\,\mathrm{\mu m}$. The corresponding longitudinal density profiles are shown in Fig~\ref{fig:Fig1}(c). This profile variation  increases the plasma wakefield in the final stage of acceleration, thus modifying the relative contribution of beam loading effects. Two simulations were performed, one with $C_\mathrm{N_2}=0.5\%$ and $d_{plate}=1000\,\mathrm{\mu m}$ and the second with $C_\mathrm{N_2}=0.35\%$ and $d_{plate}= 750 \,\mathrm{\mu m}$. For $C_\mathrm{N_2}=0.5\%$, the beam-loaded electric field is larger and a thicker plate  was used. The obtained results are presented in Table~\ref{tab:Tab.1} and compared to the ones obtained with $d_{plate}=500\,\mathrm{\mu m}$.

\begin{table}[htb]
\caption{Beam properties for $C_\mathrm{N_2}=0.5$ and $0.35\%$ and for plate aperture thickness $d_{plate}=500$, $1000\,\mathrm{\mu m}$, and cell lengths $L_{cell}=1000, 1100, 1300, 1400\,\mathrm{\mu m}$. The beam properties are average energy $\left<\mathcal{E}\right>$, rms energy spread $\Delta\mathcal{E}_{rms}/\left<\mathcal{E}\right>$ and angular divergence $\theta_{x,y}$.\label{tab:Tab.1}}

\begin{tabular}{c|c|c|c|c|c|c}
\begin{tabular}[c]{@{}c@{}}$C_\mathrm{N_2}$ \\ (\%)\end{tabular} &
\begin{tabular}[c]{@{}c@{}}$L_{cell}$ \\ ($\mu m$)\end{tabular} &
\begin{tabular}[c]{@{}c@{}}$d_{plate}$ \\ $(\mathrm{\mu m})$\end{tabular} & \begin{tabular}[c]{@{}c@{}}$\left<\mathcal{E}\right>$ \\ $(\mathrm{MeV})$\end{tabular} & \begin{tabular}[c]{@{}c@{}}$\Delta \mathcal{E}_{rms}/\left<\mathcal{E}\right>$ \\ ($\%$) \end{tabular}& \begin{tabular}[c]{@{}c@{}} $\theta_{x} $ \\ $(\mathrm{mrad})$ \end{tabular}& \begin{tabular}[c]{@{}c@{}} $\theta_{y} $ \\ $(\mathrm{mrad})$ \end{tabular} \tabularnewline
\hline
\hline
0.5 & 1000 &500 & 135 & 5.0 & 2.8 & 3.1\tabularnewline
\hline
0.5 & 1000 &1000 & 149 & 8.0 & 2.8 & 3.5\tabularnewline
\hline
0.5 & 1100 &500 & 154 & 5.4 & 2.8 & 3.2\tabularnewline
\hline
0.5 & 1400 &500 & 197 & 6.3 & 2.8 & 3.0\tabularnewline
\hline
0.35 & 1000 &500 & 145 & 4.0 & 2.3 & 3.1\tabularnewline
\hline
0.35 & 1000 &750 & 152 & 4.0 & 2.4 & 3.1\tabularnewline
\hline
0.35 & 1300 &500 & 196 & 3.2 & 2.2 & 3.1\tabularnewline
\hline
*0.5 & 1000 & 500 & 144 & 3.5 & 2.1 & 3.6\tabularnewline
\end{tabular}
\begin{flushleft}
*with laser parameters: $w= 17\,\mathrm{\mu m}$, $z_\mathrm{foc}=2.9\,\mathrm{mm}$.
\end{flushleft}
\end{table}

As shown in Table \ref{tab:Tab.1}, the average energy increases with the plate thickness for a fixed nitrogen concentration and cell length. This increase is rather small: the average energy increases by 10\% for $C_\mathrm{N_2}=0.5\%$ with $d_{plate}=1000\,\mathrm{\mu m}$ and $5\%$ for $C_\mathrm{N_2}=0.35\%$ with $d_{plate}=750\,\mathrm{\mu m}$. A larger effect is however observed on the rms energy spread especially for $C_\mathrm{N_2}=0.5\%$ at which $\Delta \mathcal{E}_{rms}/\left<\mathcal{E}\right>$ increases by $60\%$ with a thicker $d_{plate} = 1000\,\mathrm{\mu m}$. The angular divergence does not change when $d_{plate}$ is increased. For $C_\mathrm{N_2}$= 0.35\%, the properties of the high energy electron bunch are little affected by the modification of the plate  thickness.

These results can be explained by noting that the local plasma density increase induced by larger $d_{plate}$ values leads to an increase of the plasma transverse focusing force. Therefore the transverse width of the high energy electron bunch is reduced, its charge density is thus increased leading to a large amount of beam loading, which can, in the final stage of the acceleration, overcompensate the plasma wakefield. These results show that the laser-plasma configuration with $C_\mathrm{N_2}$= 0.35 \% is very stable regardless of the modification of the down ramp of the longitudinal density profile.

\subsection{Extension to higher energies by extending the plateau length}
\label{sec:extension}

As the plasma length is shorter than electron dephasing length and laser depletion length, electron energy can be increased by extending the length of the plateau density, keeping all other parameters unchanged.  Electron energies close to $200 \,\mathrm{MeV}$ can be achieved as shown in Table \ref{tab:Tab.1}, for $C_\mathrm{N_2}$= 0.5 and 0.35 \%, for $L_{cell}=1.4$ and $1.3\,\mathrm{mm}$ respectively.

The angular divergence of the beam is nearly independent of the average energy at the exit of the plasma, while a larger variation of the energy spread is observed: an increase by $26 \%$ for $C_\mathrm{N_2}=0.5\%$ but a decrease by 20\% for $C_\mathrm{N_2}=0.35 \%$. For $C_\mathrm{N_2}=0.35 \%$, the variation of $\Delta \mathcal{E}_{rms}$ over the energy range from $145$ to $196\,\mathrm{MeV}$ is only of 8\%, indicating that the acceleration is quite uniform for high energy electrons. Here again the configuration with $C_\mathrm{N_2}$= 0.35 \% is the most stable.

\section{Influence of laser parameters}
\label{sec:laser}
As shown in Sections \ref{sec:downramp} and \ref{sec:extension}, the injector configuration with $C_\mathrm{N_2}=0.35\%$  generates a high-quality electron beam. It is  robust to  modifications of plasma target parameters, and allows tuning of beam energy in the range between $150$ and $200\,\mathrm{MeV}$. In the case with $C_\mathrm{N_2}=0.5\%$, due to higher beam-loaded electric field, the electron beam properties are more sensitive to variations of plasma target parameters. Here we study the sensitivity of the generated beam properties to changes of laser parameters.

We have performed a calculation for $C_\mathrm{N_2}=0.5$\%, and a  laser waist $w$ slightly  increased from $16$ to $17 \,\mathrm{\mu m}$, and the focal position $z_{foc}$ is moved from $2.8$ to $2.9\, \mathrm{mm}$. The evolution of the normalized vector potential for these new values is plotted as a gray curve in Fig.~\ref{fig:fig2}. The beam properties obtained from this calculation are summarized in the last line of Table~\ref{tab:Tab.1}.

As can be observed in Fig.~\ref{fig:fig2}, with  these values of $w$ and $z_{foc}$, the maximum $a_0$ increases by 10\%   as compared to the previous case (red dashed-dotted line). In addition, its position is shifted towards higher $z$, and this shift is larger in the down ramp than in the up ramp of the longitudinal density profile (red solid curve). Note that there is also a slight decrease of the gradient of $a_{0}$ between $2.1\,$ and $2.5\,\mathrm{mm}$, the corresponding $a_{0}$ values are $1.5$ and $2.0$,  which are the threshold values for ionization processes $\mathrm{N^{5+}\rightarrow N^{6+}}$ and $\mathrm{N^{6+}\rightarrow N^{7+}}$ respectively. This indicates that the delay between the starting time for the injection of the $6^{th}$ and the $7^{th}$ electrons will be slightly lengthened. Using these new laser parameters, the properties of the high energy electron bunch are as follows (for each quantity we provide its value and in parenthesis its variation compared to the value obtained with previous laser parameters): average energy $\left<\mathcal{E}\right>=144\, (+ 7\%)$; charge $Q= 43\, \mathrm{pC}\, (+ 17\%)$; rms energy spread $\Delta \mathcal{E}_{rms}/\left<\mathcal{E}\right>=3.5\, (- 29\%)$; normalized emittance $\varepsilon_{x,n}=0.8\, \mathrm{mm.mrad}\, (-29\%)$, $\varepsilon_{y,n}=2.0\, (+3\%)$; rms bunch duration $\sigma_{t}=3.3\,\mathrm{fs}\, (+12\%)$; rms angular divergence $\theta_{x}=2.1\,\mathrm{mrad} (-24\%)$, $\theta_{y}=3.6\, \mathrm{mrad}\, (+15\%)$; rms apparent transverse size $\sigma_{x}=1.4\,\mathrm{\mu m}\, (-13\%)$, $\sigma_{y}=2\,\mathrm{\mu m}\,(-16\%)$; average charge density $\left<\mathrm{d}Q/\mathrm{d}\mathcal{E}\right>=2.13 \, \mathrm{pC/MeV}\, (+54\%)$; average linear charge density $Q/\sigma_{z}=43.4\, \mathrm{pC/\mu m}\, (+5\%)$.

Most  beam properties have a variation significantly larger than the $10\%$ variation of  $a_0$, due to the strongly nonlinear nature of the laser-plasma interaction process. Most of these variations are beneficial, in particular the energy spread is reduced by 29\%, while the charge is increased by 17\%. The generated beam properties with this configuration are quite similar to the best ones obtained,  with previous laser parameters for $C_\mathrm{N_2}= 0.35 \%$. The generated beam charge is $43.4\,\mathrm{pC}$, corresponding to a $60\%$ increase, and significantly higher than the $27.3 \,\mathrm{pC}$ charge obtained at $C_\mathrm{N_2}$= 0.35 \%, .

From the present study, we conclude that it is possible to find an optimized configuration with high -quality electron beam properties when increasing $C_\mathrm{N_2}$ within a limited range to increase the accelerated electron bunch charge, here at the level of $40\,\mathrm{pC}$.  However, in this optimized configuration, the performance of the injector will be very sensitive to variations of the laser and plasma parameters, because one has to achieve a precise compensation of the large beam-loaded electric field. By reducing the accelerated charge per shot by roughly a factor of two, the constrains on this compensation are greatly reduced, thus the performance of this configuration is more stable, and one can expect a significant reduction of shot-to-shot fluctuations.

\section{Conclusion}
In this article, we have investigated the influence of beam loading effects on generated beam properties in a realistic laser-plasma configuration via the control of nitrogen concentration. In optimized conditions, beam loading effects were demonstrated to be beneficial to achieve a high-quality electron bunch for accelerator application. The best results were obtained for $C_\mathrm{N_2}=0.35\%$ with an average energy $\left<\mathcal{E}\right>$ of $145\,\mathrm{MeV}$, an energy spread $\Delta\mathcal{E}_{rms}/\left<\mathcal{E}\right>$ of $4\%$, an emittance of $\varepsilon_{x,n} = 0.93\, \mathrm{mm.mrad}$ and $\varepsilon_{y,n}= 1.92\, \mathrm{mm.mrad}$, an angular divergence of $\theta_x = 2.3\,\mathrm{mrad}$ and $\theta_y = 3.1\,\mathrm{mrad}$. Except for the higher values in the $y-$direction for the emittance, due to the laser polarization effect, the generated electron bunch parameters meets the performance requirements for a laser-plasma injector.

A study of the influence of the down ramp profile shows that the plasma configuration with $C_\mathrm{N_2}=0.35\%$ generates electron beams with stable properties with regard to the down ramp profile variations.

To generate higher energy electron bunch, the cell length can be extended. For the optimal nitrogen concentration, here $C_\mathrm{N_2}=0.35\%$,  the energy of the electron bunch can be tuned up to $200\,\mathrm{MeV}$ while preserving, and even improving, other beam properties.

The sensitivity of the bunch properties to laser parameters  was studied. Due to the high beam-loaded electric field for the case with $C_\mathrm{N_2} = 0.5\%$, the electron beam properties are sensitive to  variations of the plasma properties. By adjusting laser parameters, it is also possible  to find an optimized configuration for this concentration, with a high-quality electron bunch and high charge. However, for higher $C_\mathrm{N_2}$,  generating larger electron bunch charge, the performance of the injector will become more sensitive to variations of laser-plasma parameters because the compensation of the large beam-loaded electric field has to be more precise.
These results provide fundamental  insight on the stability of laser plasma injectors.

\section*{Acknowledgments}
This work was granted access to the HPC resources of [TGCC/CINES] under the allocation 2017- [A0010510062] made by GENCI. We also acknowledge the use of the computing center MésoLUM of the LUMAT research federation (FR LUMAT 2764). T.L. Audet acknowledges financial support of EuPRAXIA, co-funded by the European Commission in its Horizon2020 Programme under
the Grant Agreement no 653782.
\bibliographystyle{apsrev4-1}
\bibliography{main}

\end{document}